\documentclass[sigconf,natbib=true]{acmart}
\AtBeginDocument{%
  }

\copyrightyear{2023}
\acmYear{2023}
\setcopyright{acmlicensed}\acmConference[RecSys '23]{Seventeenth ACM Conference on Recommender Systems}{September 18--22, 2023}{Singapore, Singapore}
\acmBooktitle{Seventeenth ACM Conference on Recommender Systems (RecSys '23), September 18--22, 2023, Singapore, Singapore}
\acmPrice{15.00}
\acmDOI{10.1145/3604915.3609490}
\acmISBN{979-8-4007-0241-9/23/09}

\begin{document}

\title{RecAD: Towards A Unified Library for Recommender Attack and Defense}


\author{Changsheng Wang}
\authornote{The first two authors contributed equally to this research.}
\email{wcsa23187@gmail.com}
\orcid{0009-0007-0957-638X}
\affiliation{%
  \institution{University of Science and Technology of China}
  \streetaddress{No.96 Jinzhai Road}
  \city{Hefei}
  \state{Anhui}
  \country{China}
  \postcode{230000}
}

\author{Jianbai Ye}
\authornotemark[1]
\orcid{0009-0004-9095-4947}
\email{jianbaiye1999@gmail.com}
\affiliation{%
  \institution{University of Science and Technology of China}
  \city{Hefei}
  \state{Anhui}
  \country{China}
}

\author{Wenjie Wang}
\email{wenjiewang96@gmail.com}
\orcid{0000-0002-5199-1428}
\authornote{Corresponding author.}
\affiliation{%
  \institution{National University of Singapore}
  \country{Singapore}
}

\author{Chongming Gao}
\email{chongming.gao@gmail.com}
\orcid{0000-0002-5187-9196}
\affiliation{%
  \institution{University of Science and Technology of China}
  \streetaddress{No.96 Jinzhai Road}
  \city{Hefei}
  \state{Anhui}
  \country{China}
  \postcode{230000}
}

\author{Fuli Feng}
\email{fulifeng93@gmail.com}
\orcid{0000-0002-5828-9842}
\affiliation{%
  \institution{University of Science and Technology of China}
  \streetaddress{No.96 Jinzhai Road}
  \city{Hefei}
  \state{Anhui}
  \country{China}
  \postcode{230000}
}

\author{Xiangnan He}
\email{xiangnanhe@gmail.com}
\orcid{0000-0001-8472-7992}
\affiliation{%
  \institution{University of Science and Technology of China}
  \streetaddress{No.96 Jinzhai Road}
  \city{Hefei}
  \state{Anhui}
  \country{China}
  \postcode{230000}
}

\renewcommand{\shortauthors}{Wang et al.}

\begin{abstract}
  In recent years, recommender systems have become a ubiquitous part of our daily lives, while they suffer from a high risk of being attacked due to the growing commercial and social values. 
  Despite significant research progress in recommender attack and defense, there is a lack of a widely-recognized benchmarking standard in the field, leading to unfair performance comparison and limited credibility of experiments. 
  To address this, we propose RecAD, a unified library aiming at establishing an open benchmark for recommender attack and defense. 
  RecAD takes an initial step to set up a unified benchmarking pipeline for reproducible research by integrating diverse datasets, standard source codes, hyper-parameter settings, running logs, attack knowledge, attack budget, and evaluation results. 
  The benchmark is designed to be comprehensive and sustainable, covering both attack, defense, and evaluation tasks, enabling more researchers to easily follow and contribute to this promising field. 
    RecAD will drive more solid and reproducible research on recommender systems attack and defense, reduce the redundant efforts of researchers, and ultimately increase the credibility and practical value of recommender attack and defense. 
    The project is released at 
    \url{https://github.com/gusye1234/recad}.
\end{abstract}

\ccsdesc[500]{Information systems~Recommender systems}

\keywords{Recommender Systems; Shilling Attack and Defense; Benchmark}

\maketitle

\section{Introduction}

In recent decades, recommender systems have become increasingly important in various areas, including E-commerce recommendation \cite{4338497}, short video entertainment \cite{gao2022cirs}, news headlines \cite{10.1145/1719970.1719976}, and online education \cite{wu2011semi}. However, the widespread use of recommender systems has also led to concerns regarding their security. When an attacker successfully attack a recommender system, users may be offended by malicious recommendations, and the platform owner may lose the trust of users in the platform. Additionally, merchants who rely on recommender platforms may suffer from commercial losses. Malicious attacks may even cause recommender systems to engage in unethical or illegal behavior, resulting in adverse effects on society. Therefore, it is essential to address these security risks and ensure the integrity of recommender systems to maintain user trust and prevent any negative consequences. 

Recently, industry and academia are trying to develop strategies for both attacking and defending recommender systems, especially pay attention to the techniques about shilling attacks and defense \cite{182952}. In shilling attacks, fake users are generated and assigned high ratings for a target item, while also rating other items to act like normal users for evading. 
There are three main kinds of shilling attack methods: heuristic methods, gradient methods, and neural methods. Heuristic methods \cite{1167344,burke2005limited,kaur2016shilling} involve artificially or randomly selecting items based on user preferences, and then intuitively fabricate interaction information. Gradient methods \cite{li2016data,fang2020influence} estimate the gradients of maximizing attack objectives to directly optimize the interactions of fake users, while Neural methods \cite{tang2020revisiting,Lin_2022,lin2020attacking} optimize the neural networks parameters to predict the optimal fake user behaviors for maximizing attack objectives.

To combat these attacks, many defense methods \cite{dou2018collaborative,cao2013shilling,mehta2009unsupervised} are emerging to enhance the defense ability of existing recommendation models. Essentially, these defense models aim to distinguish between fake data generated by the attack model and real user data, ensuring that the recommendation model uses as much real data as possible \cite{yang2018detection}. Currently, the mainstream defense model can be divided into three types according to whether the label of fake users can be obtained \cite{dou2018collaborative,aktukmak2019quick,tang2019adversarial}. As new attack methods emerge, defense models are constantly evolving to keep up with these threats. Therefore, staying up-to-date with the latest attack and defense methods is essential for maintaining the security of recommender systems.

With the continuous emergence of new attack and defense algorithms in the field of recommender system security, there are several research challenges that deserve attention. 
Firstly, while many articles provide details about their experiments, there is often a lack of standardization in dataset processing methods, which can lead to unfair comparison. Secondly, there is a lack of unified settings for attack experiments. Various works usually leverage different experimental setting, making it difficult to compare and evaluate different models. It is critical to establish a standardized approach for similar attack settings to facilitate the comparison and evaluation of different models. Additionally, many works lack public code, which can create repetition and difficulties for subsequent researchers trying to advance the field. 
To address these challenges, researchers should strive to provide clear and standardized descriptions of dataset processing methods, unified settings for attack experiments, and make their code publicly available to facilitate replication and extension of their work. These efforts can help promote the development of the recommender attack and defense and contribute to more robust and effective recommender system security solutions.

To address the aforementioned challenges, we have initiated a project to develop a unified framework for \underline{Rec}ommender \underline{A}ttack and \underline{D}efense, named RecAD. RecAD aims to improve the reproducibility of existing models and simplify the development process for new recommender attack and defense 
 models. 
Our benchmarking library design is innovative and effective, revealing several advantages compared to earlier attempts.
\begin{itemize}
    \item \textbf{Unified library framework.}
    RecAD is implemented using Pytorch\footnote{\url{https://pytorch.org/}.}, one of the most popular deep learning frameworks. The library is composed of three core modules, namely the data module, model module, and evaluation module. The library supports a vast array of options provided by each module, and a straightforward configuration ensures that users can promptly complete algorithm reproduction and comparison. The seamless interface integration of the three core modules also enables the minimal adjustment for incorporating new algorithms, allowing for continuous development and extension within our framework in the future.
    \item \textbf{Comprehensive benchmark models and datasets.}
    RecAD provides support not only for replacing individual models but also for integrating a wide range of research issues. From generating fake attack data to defending against existing data and injecting data into victim models, RecAD covers the entire spectrum of shilling attack and defense research. It provides an array of choices for all models and datasets, guaranteeing an ample assortment of combinations for researchers to utilize. This allows them to execute, compare, and assess the entire procedure, relying on lucid instructions and configurations. RecAD is highly adaptable and scalable, with original dataset copies that can be effortlessly transformed into a practical form using the provided preprocessing tools or scripts. Additionally, we are continuously expanding our library with additional datasets and methods to better serve the needs of the community and researchers.
    
    \item \textbf{Extensive and standard evaluation protocols. } 
    RecAD offers evaluation methods from two perspectives: attack evaluation and defense evaluation. Researchers interested in continuing the offensive direction or those focusing on the defensive direction can use the corresponding evaluation methods. Additionally, it provides standard evaluation techniques for assessing the effectiveness of defense models, encapsulating the entire evaluation process within a singular module enables RecAD to more readily accommodate more evaluation techniques, thus enhancing its adaptability and versatility.


    
    \item \textbf{Openness and high integration of models. }
    Openness is crucial for promoting transparency, collaboration, and reproducibility in computer science research. 
    RecAD adopts a highly integrated approach, simplifying the relationships between modules as much as possible and making the corresponding parameters publicly available at each module. This ensures that subsequent researchers who use our framework to add new models only need to make the corresponding modules public, allowing other researchers to quickly and efficiently reproduce the work and ensure the openness of the field in the future.
    
    \item \textbf{The generalization of attacker's knowledge. }
    The attacker's knowledge level directly impacts the effectiveness of the attack. A high degree of accessible knowledge about the recommender system allows an attacker to craft adversarial examples that can evade the model's defenses. RecAD can elevate white-box attacks to gray-box attacks and customize the proportion of data accessible by the attackers for gray-box attacks \cite{10.1145/3274694.3274706}, promoting the fair comparison between a wide range of attackers.  
\end{itemize}

\section{Related Work}
\subsection{Overview of Shilling Attack and Defense}
In the past two decades, researchers have conducted experiments to demonstrate the feasibility of attacking real-world recommender systems, such as YouTube, Google Search, Amazon, and Yelp. These experiments have shown that it is possible to manipulate recommendation systems in practice, resulting in an increasing focus on this field from both the academic community and industry. To promote its development, researchers have typically focused on either shilling attacks or defense mechanisms.
With the advancements in deep learning, the field has seen a notable increase in the effectiveness of these methods.


\subsection{Shilling attack}

%
The objective of an shilling attack is to interfere with the recommendation strategy of the victim recommender system through a series of measures \cite{o2005recommender,mobasher2007toward,deldjoo2019assessing}. The ultimate goal is to enhance the exposure of a specific target item among all users after the recommender model is trained.
To achieve this objective, attackers often inject fake users into the historical interactive data, or training matrix, of the recommender system. However, if these fake users are not adequately protected, they will be sent into the recommender system model during the training process, thus disrupting the recommendation strategy of the system. As a result, the key challenge of an shilling attack is to construct the interaction behaviors of the fake users.
The interaction behaviors of the constructed users can generally be classified into three categories:
\begin{itemize}
    \item \textbf{Heuristic attacks. }
Heuristic attacks involve selecting items to create fake profiles based on subjective inference and existing knowledge. The goal is to strengthen the connection between fake users and other real users while evading defense methods and achieving exposure enhancement of the final target item \cite{1167344,burke2005limited}. Currently, existing methods include the Random Attack \cite{kaur2016shilling}, Average Attack \cite{lam2004shilling}, Bandwagon Attack, and Segment Attack \cite{1565730}. 
The Random Attack is a low-knowledge method that selects filler items randomly, while the Average Attack selects filler items randomly and requires more knowledge. In the case of an Average Attack, the target item needs to be given the highest rating to implement a push attack. Segment Attack selects items of the same category as the target item and maximizes their rating, with the goal of creating a stronger correlation with the corresponding target user among real users so that it can attack more effectively.
    \item \textbf{Gradient attacks.}
Gradient attacks involve relaxing the discrete space to a continuous space to ensure that the objective function can be optimized by the gradient to achieve the optimal attack effect. For instance, Li et al. \cite{li2016data,fang2020influence} developed poisoning attacks optimized for matrix factorization-based recommender systems, while \citet{yang2017fake} developed poisoning attacks optimized for co-visitation rule-based recommender systems. Additionally, there are gradient attack methods based on Next-item \cite{zhang2020practical}, and graph \cite{fang2018poisoning}.
However, all Gradient Attacks require known types of recommender systems to carry out specific optimization, which does not have good generalization. Moreover, in order to achieve bi-level optimization \cite{huang2021data}, directly adjusting interactions according to gradients involved transforming the discrete interactions into continuous optimization. During the process of re-discretization, information loss occurred, leading to sub-optimal results and the lack of robustness in the model.
    \item \textbf{Neural attacks. }
Neural Attacks, primarily inspired by deep learning \cite{huang2021data}, generate realistic profiles that have a significant impact on recommender systems by optimizing the parameters of neural networks to maximize the objective function. WGAN \cite{https://doi.org/10.48550/arxiv.1701.07875} draws on Wasserstein's distance, which has shown better empirical performance than the original GAN \cite{goodfellow2020generative}. It can emulate real user behavior with fake user behavior to achieve the effect of fake user behavior.
AIA \cite{tang2020revisiting} reviewed the bi-level optimization problems of the surrogate model and proposed time-efficient and resource-efficient solutions. AUSH \cite{lin2020attacking} and Legup \cite{Lin_2022} solve the randomness caused by noise generation in common models, making the generated template artificially based on known knowledge, resulting in a more undetectable configuration file.
When the attacker's knowledge is limited to a black box, researchers use RL attack \cite{zhang2022targeted,fan2021attacking,song2020poisonrec} to complete the attack, with the attacker adjusting changes based on feedback given by the spy user in the victim model. The methods of Neural Attacks all show better performance on real datasets than Gradient and Heuristic attacks.
\end{itemize}

In addition to the challenges associated with constructing effective shilling attacks, another emerging issue is the \textbf{knowledge of the attacker} \cite{burke2005limited}. In today's world, data security and privacy are increasingly important to both users and companies. This makes it increasingly challenging for attackers to obtain the necessary user data to construct effective attack models.
As a result, researchers have begun to consider the attacker's knowledge as a key constraint for the attack model. The attacker's knowledge can be classified into three categories: \emph{white box}, \emph{grey box}, and \emph{black box}. In a white box attack, the attacker has complete knowledge of the target recommender model, which includes all the data of the victim model used for training and the network structure and parameters of the victim model. In a grey box attack, the attacker can only access part of the training set of the target model and has no knowledge of the victim model. In a black box attack, only some spy users are allowed as attack feedback.

In addition to shilling attacks, there are other types of attacks, such as attacks that involve modifying the real user interaction history \cite{zeller2008cross} or attacks based on federated learning recommender models \cite{rong2022fedrecattack,rong2022poisoning,zhou2021deep,https://doi.org/10.48550/arxiv.2202.06701}. The former is not very effective due to the adoption of multiple privacy protection mechanisms by real recommender platforms, such as email and mobile phone hardware binding. Therefore, this method is easily detected and defended by the platform and is insufficient for a large-scale attack.
On the other hand, the latter is still in the theoretical research stage and the models proposed are too basic, at the same time this kind of method has not yet been implemented by companies. This means that the criteria for the above two methods still need to be explored by more researchers.
\subsection{Defense}
A defense model can be viewed as a checkpoint responsible for detecting possible fake users in the data before it is sent to the recommender model. The defense model eliminates fake users to ensure that the recommendation results are not interfered with by attackers to the greatest extent possible. Some defense models attempt to find the law of data distribution from all the data or obtain the probability of the corresponding label through probability methods to predict and classify.
Currently, the defense direction can be classified into three categories:
\begin{itemize}
    \item \textbf{Supervised defense models, }
which need to be pre-labeled with true and false data. The goal of the model is to learn the relationship between the input and output variables, so that it can make predictions on new data. The learning process involves minimizing the difference between the predicted output and the true output for each example in the training data. In other words, the model is trained to approximate the mapping from inputs to outputs. In the direction of recommender defense, Supervised work emerges in the initial exploration of this field, such as CoDetector \cite{dou2018collaborative}, DegreeSAD \cite{li2016shilling}, BayesDetector \cite{yang2018detection}.
    \item \textbf{Semi-supervised defense models, }
as explored in \cite{cao2013shilling, wu2011semi}, aim to use a minimal amount of false data while still maintaining the purpose and accuracy of the supervised method. This is because attackers typically use a small amount of data to launch attacks, leading to an inherent imbalance between true and false training samples, highlighting the crucial importance of maintaining the supervised aspect of the method.
    \item \textbf{Unsupervised defense models, }
which have been intensively investigated in recent years, including traditional machine learning models such as probabilistic models \cite{aktukmak2019quick}, statistical anomaly detection \cite{bhaumik2006securing}, PCA \cite{mehta2009unsupervised}, SVM \cite{zhou2016svm}, and K-means \cite{davoudi2017detection}. More recently, network models have been used for detection, such as Graph Embedding \cite{10.1145/3442381.3449813}, Sequential GANs \cite{DBLP:journals/corr/abs-2012-02509}, Recurrent Neural Network \cite{gao2020shilling}, and Dual-input CNN \cite{9498095}.
\end{itemize}

\smallskip
In addition to the model-based prediction introduced above to realize the defense of the recommender platform, some scholars also have trained the recommender model by using adversarial data training \cite{tang2019adversarial, he2018adversarial, liu2020certifiable, wu2021fight} so that the recommender model can have better generalization in the face of fake data.

\subsection{Benchmarking for Recommender Attack and Defense}

Despite the recent growth in the field of RS security, different studies have employed different data sets, evaluation methods, and knowledge constraints, resulting in significant fairness issues when comparing different models. This has had a negative impact on the steady development of the field. Although some works have attempted to address these issues in the past, there is still a need for a comprehensive and unified library to solve the current dilemma.

For instance, in AUSH \cite{lin2020attacking}, the author provided a code that integrated multiple attack models, but the workflow was inefficient and required a significant amount of time for subsequent researchers to study the code structure. Additionally, the code was not friendly for adding new models under the same framework and focused more on the study of attack models. Moreover, the code only provided a limited data set and did not include the data processing method, making it difficult to test the model on a working public data set. In SDLib\footnote{\url{https://github.com/Coder-Yu/SDLib}.}, some defense models and attack models were provided, but the attack model was outdated and did not complete the entire process from attack generation to defense detection and injection into the recommender model. Furthermore, the code language used in this work was obsolete. Our framework overcomes the limitations of previous methods by abstracting each component into relatively independent modules, ensuring the unity and extensibility of the model. This allows for better maintenance and development of the framework in the future.

\begin{figure}[tbp]
  \centering
  \includegraphics[scale=0.55]{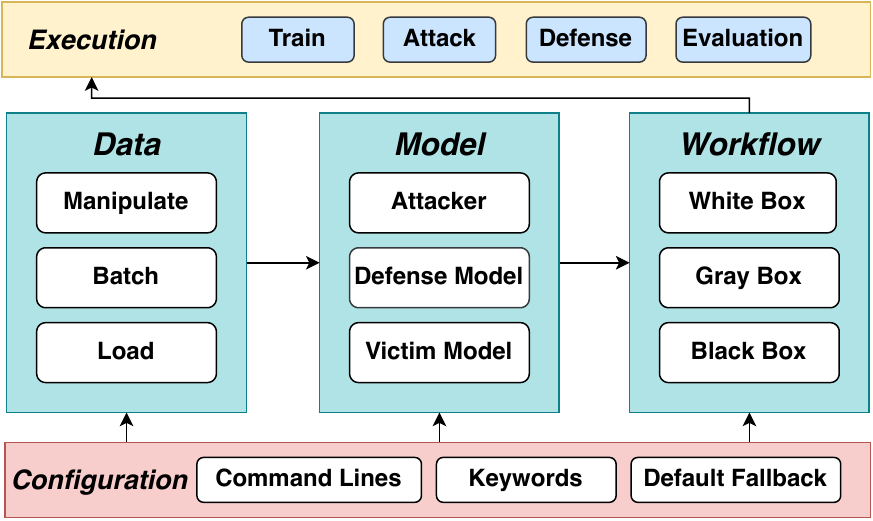}
  \caption{
  The overall framework of RecAD.
  }
  \label{fig:1}
\end{figure}

\begin{table}[tbp]
\centering

\caption{Collected data in our library.}
\label{table:1}
\begin{tabular}{l|cccc}
\hline
Dataset                            & \#Users                     & \#Items                     & \#Interations     & \#Density                  \\ \hline \hline
MovieLens-1m*                      & 5,950                       & 3,702                       & 567,533         & 0.257\%                \\
Yelp*                              & 54,632                      & 34,474                      & 1,334,942       & 0.070\%                \\
Amazon-Game*                       & 3,179                       & 5,600                       & 38, 596         & 0.216\%                 \\
Book-Crossing                      & 105,284                     & 340,557                     & 1,149,780       & 0.003\%                 \\
Last.FM                            & 1,892                       & 17,632                      & 92,834          & 0.278\%                 \\
Epinions                           & 116,260                     & 41,269                      & 188,478         & 0.004\%                 \\
Gowalla                            & 107,092                     & 1,280,969                   & 6,442,892      & 0.005\%                  \\ 
\hline
\multicolumn{5}{l}{*\footnotesize{means the dataset is used in the experiments and only kept high-frequency users}} \\[1px]
\multicolumn{5}{l}{\footnotesize{\: and items (at least 10 interactions).}} \\
\end{tabular}
\end{table}

\footnotetext[2]{\url{https://grouplens.org/datasets/movielens/1m/}.}
\footnotetext[3]{\url{https://www.yelp.com/dataset}.}
\section{The Library-RecAD}\label{sec:problem}

The overall framework of RecAD is illustrated in Figure \ref{fig:1}. At the bottom, our library maintains a flat structure for the default hyper-parameters globally, and the core components are built upon it with automated parameter loading (see Section \ref{sec:method}). Our library abstracts the core modules at three levels: data, model, and workflow. 
In the following, we briefly present the designs of these three core modules.

\begin{figure*}[]
  \centering
  \includegraphics[width=\textwidth]{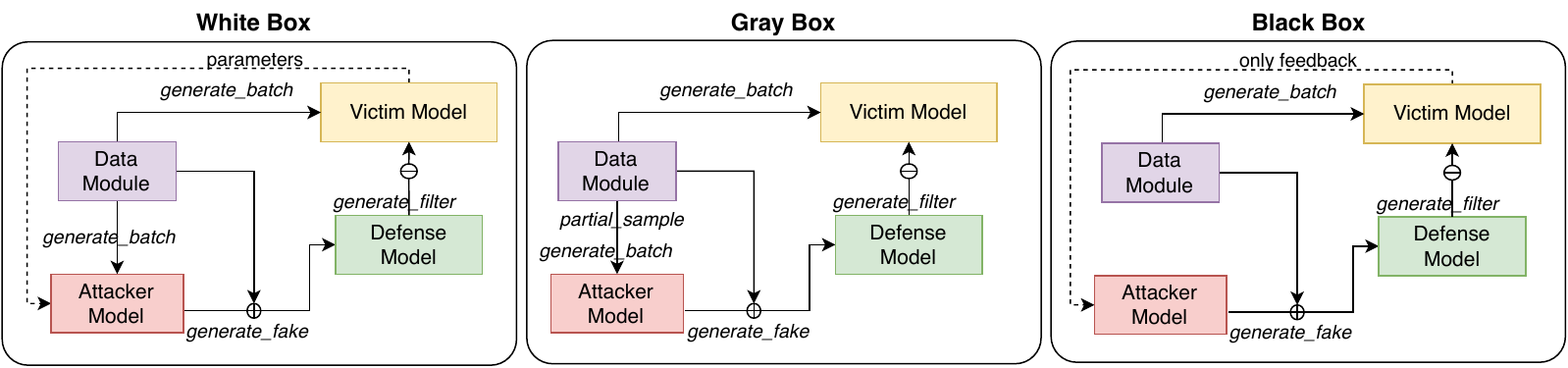}
  \caption{
  Component workflow under different attack knowledge.
  }
  \label{fig:3}
\end{figure*}



\subsection{Data Module}
The data module serves as the fundamental part of the entire library, as it provides essential runtime information such as batches and indicators of scale. It takes charge of dataset loading, batch generation, and fake data manipulation.
\subsubsection{Dataset Loading}
To create an actively-contributed benchmark, it is important to make the addition of new datasets as easy as possible. Therefore, we have designed the data module to keep the required dataset formats simple and flexible. At present, our library only requires the human-readable CSV format with specific column names to load datasets into explicit or implicit interactions. This design decision allows users to easily add their own datasets to the library without having to modify the codebase. Our library already supports multiple datasets (as shown in Table \ref{table:1}), and we also provide auxiliary functions to convert datasets from other well-known recommender frameworks, such as RecBole \cite{recbole[1.0]}. This provides further flexibility for users to utilize the datasets that they are familiar with.

\subsubsection{Batch Generation}
Our library prioritizes seamless integration between datasets, models, and workflows, which presents challenges for batch generation. 
To address this, we design a flexible and generic interface (\textit{generate\_batch}).
The interface allows the caller to provide runtime configuration parameters (\emph{e.g.}, pairwise sampling; binarizing the ratings) and dispatches itself to the corresponding behavior.
This design reduces the workloads on developers who are attempting to adapt their data and allows them to focus on providing as much runtime information as possible.

\subsubsection{Fake Data Manipulation}
In our library, we recognize the importance of addressing the manipulation of fake data during runtime. 
Specifically, we must account for both the injection of fake data from attacker models and the filtering of fake data by defense models.
We address this challenge with unified interfaces named \textit{inject\_data} and \textit{filter\_data}, respectively. These interfaces are called by the attacker and defense models to manipulate the dataset.

\begin{figure}[tbp]
  \setlength{\belowcaptionskip}{-0.2cm}
  \centering
  \includegraphics[width=0.45\textwidth]{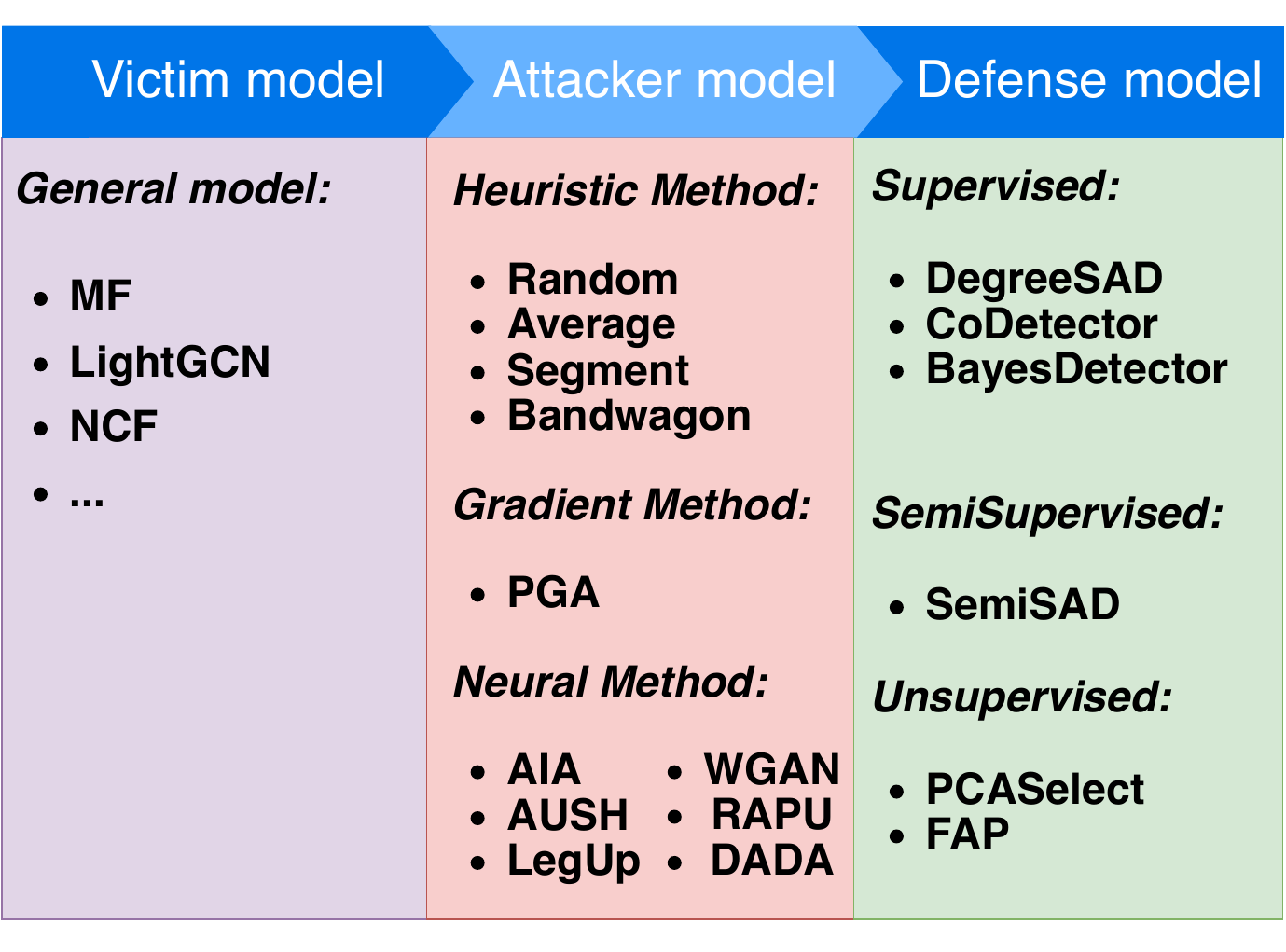}
  \caption{
  The models that are supported by RecAD.
  }
  \label{fig:2}
\end{figure}

\subsection{Model Module}
The model implementation is the most versatile part of the library, and we offer maximum flexibility to accommodate different approaches. 
To account for the similarities and differences between models, we introduce a general base model and its successors: the victim, attacker, and defense models. Figure \ref{fig:2} presents the models that have been implemented.

\subsubsection{Base Model} 
We don't provide framework-level abstractions for model optimization. 
Instead, the models are responsible for their own single-epoch training and evaluation, which can be implemented through a set of auxiliary functions provided by the library. 
This design choice is aimed at reducing the complexity of the framework and enabling the integration of a wide range of models, without requiring modification of the framework-level abstractions for each individual model. 
 To facilitate this, we use unified interfaces (\textit{train\_step}, \textit{test\_step}) that enable the callers to initiate the training or evaluation process of the models. 

\subsubsection{Victim Model}
Victim models are recommender models, and the library provides a unified interface for training and testing them. This makes it easy to integrate any victim model into the library without the need for modification of the core framework.

\subsubsection{Attacker Model}
In our library, the training of the attacker model shares the same interface with victim models (\emph{i.e.} \textit{train\_step}).
After training, the attacker model generates the fake data through a unified interface (\textit{generate\_fake}) and then forwards the contaminated data to the next module.
Since the full set of the dataset is not necessarily exposed (\emph{e.g.} Gray box attacking in Figure \ref{fig:3}), \textit{generate\_fake} should explicitly receive the target dataset as a parameter.

\subsubsection{Defense Model}
The defense model is trained on the attacked data through the same training interface. 
The objective of the model is to output a filtered dataset with fake data removed.
Our library summarizes a unified interface \textit{generate\_filter} to wrap the implementation details of each defense model.

\subsection{Workflow Module}
This module is the corresponding abstraction of different attack knowledge (Figure \ref{fig:3}). 
The workflow module holds the instantiations of the data module and model module, controlling the exposure of data and the interaction of modules. 
It also contains the boilerplate code for the training loop and evaluation callbacks (\emph{e.g.}, early stop; report after training).
\subsubsection{Data Exposure}
The data exposure level for different models varies depending on the attack knowledge settings and the running stages (as shown in Figure \ref{fig:3}). For instance, the attacker model may be exposed separately to full, partial, or zero training data. 
Similarly, the victim model may be trained on clean data initially and later re-trained on the contaminated data during the attack process. 
The workflow module in our library is responsible for constructing the appropriate data flow according to the attack knowledge and ensuring that no accidental data leakage occurs.
This way, our library provides a flexible and secure environment for implementing and testing various attack and defense models under different settings.

\subsubsection{Module Interaction}
The interactions between modules vary between attacks.
In a white-box attack, the model has direct access to all the training data, whereas, in a black-box attack, the model receives feedback from the victim without any access to the training data. 
For workflows \cite{tang2020revisiting, Lin_2022} where no defense model is involved, the fake data generated by the attacker model flows directly into the victim's training without filtering. 
The workflow module arranges the dependencies of modules and prevents any inappropriate interactions between them.

\subsubsection{Training \& Evaluation}
In order to better control the data exposure and module interaction, we give the workflow module the responsibility for launching the training and evaluation of the contained models. 
The workflow module contains the boilerplate codes for wrapping the training loop outside the models' \textit{train\_step}.
Also, we design a hooking mechanism to provide flexibility for models to set up their evaluation callbacks.
This allows models to define their own evaluation metrics to evaluate the model's performance at different stages of the training process.

\begin{figure}[tbp]
\setlength{\belowcaptionskip}{-0.5cm}
  \centering
  \includegraphics[width=0.45\textwidth]{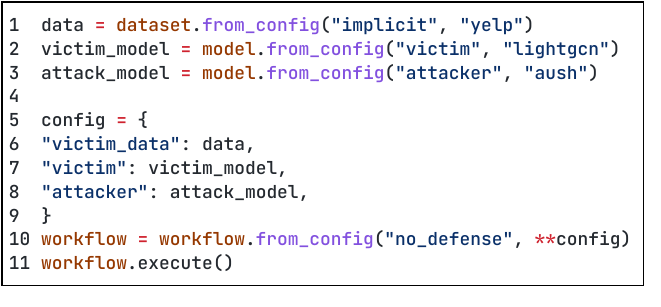}
  \caption{
  A code snippet of module instantiations of RecAD.
  }
  \label{fig:4}
\end{figure}

\begin{figure}[tbp]
  \centering
  \setlength{\belowcaptionskip}{-0.4cm}
\includegraphics[width=0.4\textwidth]{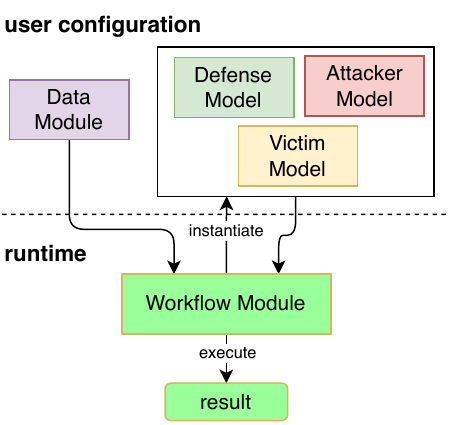}
  \caption{
  An illustration of lazy instantiation.
  }
  \label{fig:lazy}
\end{figure}

\section{USAGE GUIDELINE OF THE LIBRARY}\label{sec:method}
In the following two parts, we first show the typical usage to instantiate the existing modules of our library, 
then detail the steps to extend our library with a new implementation.

\subsection{Module Instantiations}
Attacking a recommender system often involves using multiple datasets and machine learning models, which makes the training and testing process more complex than for regular recommender systems. 
Our library simplifies this process by exposing the necessary modules to users and providing a unified interface called \emph{from\_config} for instantiating them (Figure \ref{fig:4}). 
Two kinds of parameters may be needed from \emph{from\_config}: hyper-parameters and runtime parameters. 

\subsubsection{Hyper-parameters}
Our library employs the hashing table to store all default hyper-parameters of modules together and offer global access across programs.
While instantiating, our library automatically loads default parameters on-fly from the hashing table and updates them from the keyword arguments passed by the user.
The decoupling of default hyper-parameters and the actual module implementation facilitates a quick overview of configurable parameters for the user.

\subsubsection{Runtime Parameters}
Runtime parameters are the parameters that won't be settled before the runtime. 
For example, the model module in Figure \ref{fig:4} normally needs the numbers of the user and item to create the embeddings when instantiating. 
Due to the data injection or data filtering from the attacker model, the actual numbers of the user and item are not known before the runtime. 
But the dependency between the model and data module is clear, and it is burdensome to ask the user to manually pass in the required instances in the program.
Hence, we implement lazy instantiation (Figure \ref{fig:lazy}) to make runtime parameters transparent at the user level. The module won't actually instantiate at the time the user call \emph{from\_config} if the needed runtime parameters are not passed.
Instead, the workflow will sort out the dependencies between modules and automatically fill in the required runtime parameters to complete the instantiation. This decouples the instantiation of modules from the availability of runtime parameters, making the library more flexible and adaptable to different scenarios.

\subsection{Module Extension}
In our library, we provide the base class for all the core modules: \textit{BaseData}, \textit{BaseModel}, and \textit{BaseWorkflow}. 
We require the extended module must be the corresponding base class's subclass so that the necessary abstract interfaces can be called properly.

\begin{table*}[t]
    \renewcommand{\arraystretch}{1.5}
\setlength{\abovecaptionskip}{0cm}
\setlength{\belowcaptionskip}{0cm}
\caption{Overall attack performance on three recommendation datasets.}
\label{tab:overall_per_1}
\begin{center}
\setlength{\tabcolsep}{0.6mm}{
\resizebox{\textwidth}{!}{
\begin{tabular}{cccccccccccccc}
\hline
\multicolumn{1}{l}{}                 & \multicolumn{1}{l}{}                    & \multicolumn{4}{c}{ML-1M}                              & \multicolumn{4}{c}{Yelp}                               & \multicolumn{4}{c}{Amazon}        \\
Attack Method                        & Attack Knowledge                        & HR@10  & HR@20  & HR@50  & HR@100                      & HR@10  & HR@20  & HR@50  & HR@100                      & HR@10  & HR@20  & HR@50  & HR@100 \\ \hline
\multicolumn{1}{c|}{No Attacker}     & \multicolumn{1}{c|}{None}               & 0.0050 & 0.0109 & 0.0297 & \multicolumn{1}{c|}{0.0656} & 0.0114 & 0.0190 & 0.0375 & \multicolumn{1}{c|}{0.0630} & 0.0000 & 0.0000 & 0.0003 & 0.0016 \\ \hline
\multicolumn{1}{c|}{RandomAttacker} & \multicolumn{1}{c|}{White Box}          & 0.0050 & 0.0082 & 0.0228 & \multicolumn{1}{c|}{0.0457} & 0.0078 & 0.0112 & 0.0214 & \multicolumn{1}{c|}{0.0362} & 0.0000 & 0.0000 & 0.0000 & 0.0000 \\
\multicolumn{1}{c|}{SegmentAttack}   & \multicolumn{1}{c|}{White Box}          & 0.0069 & 0.0123 & 0.0288 & \multicolumn{1}{c|}{0.0630} & 0.0057 & 0.0083 & 0.0153 & \multicolumn{1}{c|}{0.0258} & 0.0397 & 0.0520 & 0.0675 & 0.0832 \\
\multicolumn{1}{c|}{BandwagonAttack}  & \multicolumn{1}{c|}{White Box}          & 0.0059 & 0.0119 & 0.0267 & \multicolumn{1}{c|}{0.0592} & 0.0066 & 0.0114 & 0.0257 & \multicolumn{1}{c|}{0.0431} & 0.0050 & 0.0205 & 0.0523 & 0.0854 \\
\multicolumn{1}{c|}{AverageAttack}   & \multicolumn{1}{c|}{White Box}          & 0.0016 & 0.0044 & 0.0167 & \multicolumn{1}{c|}{0.0400} & 0.0053 & 0.0090 & 0.0169 & \multicolumn{1}{c|}{0.0284} & 0.0085 & 0.0170 & 0.0463 & 0.0914 \\
\multicolumn{1}{c|}{WGAN}            & \multicolumn{1}{c|}{White Box}          & 0.0023 & 0.0060 & 0.0149 & \multicolumn{1}{c|}{0.0340} & 0.0143 & 0.0177 & 0.0254 & \multicolumn{1}{c|}{0.0344} & 0.1646 & 0.1788 & 0.2043 & 0.2226 \\
\multicolumn{1}{c|}{AIA}             & \multicolumn{1}{c|}{Gray Box 20\% data} & 0.0078 & 0.0180 & 0.0459 & \multicolumn{1}{c|}{0.1007} & 0.0187 & 0.0273 & 0.0465 & \multicolumn{1}{c|}{0.0686} & 0.0441 & 0.0873 & 0.4278 & 0.4839 \\
\multicolumn{1}{c|}{AUSH}            & \multicolumn{1}{c|}{Gray Box 20\% data} & 0.0071 & 0.0151 & 0.0434 & \multicolumn{1}{c|}{0.0945} & 0.0135 & 0.0217 & 0.0393 & \multicolumn{1}{c|}{0.0617} & 0.0583 & 0.1170 & 0.4392 & 0.4805 \\
\multicolumn{1}{c|}{Legup}           & \multicolumn{1}{c|}{Gray Box 20\% data} & 0.0094 & 0.0130 & 0.0283 & \multicolumn{1}{c|}{0.0471} & 0.0068 & 0.0099 & 0.0162 & \multicolumn{1}{c|}{0.0242} & 0.1847 & 0.2015 & 0.2286 & 0.2566 \\ \hline
\end{tabular}
}}
\end{center}
\end{table*}

\subsubsection{General Module}
Two abstract methods must be implemented for all the modules:
\begin{itemize}
    \item \textbf{\textit{from\_config}}: users pass arguments to this method to instantiate a new module. Our library has already implemented the argument sanity checking and overwriting the default hyper-parameters in the father class. A new module should assign the default hyper-parameters in this method.
    \item \textbf{\textit{info\_describe}}: modules interact through this method. The method should return a hash table with the named variable that this module can expose publicly.
\end{itemize}
\subsubsection{Core Modules}
The core modules have specialized interfaces that need to be implemented in addition. We have discussed most of the below interfaces in Section \ref{sec:problem}.
\begin{itemize}
    \item \textbf{Data Module}: The most important interface for this module is \textit{generate\_batch}. The interface should take the caller's keyword arguments as the input, and return the correct batches of the dataset for later training or testing.
    \item \textbf{Model Module}: Right now, three kinds of models are considered: victim model, attacker model, and defense model. They are all required to be implemented with two interfaces: \textit{train\_step} and \textit{test\_step} to perform one-epoch training or testing. Besides, for the attacker model and defense model, \textit{generate\_fake} and \textit{generate\_filter} need to be implemented, respectively. 
    \item \textbf{Workflow Module}: An interface named \textit{execute} should be implemented for users to explicitly launch the whole workflow. Inside the interface, the implementor should correctly instantiate and arrange the modules.
\end{itemize}

\section{Experiments}

This section showcases the application of RecAD by implementing various representative attackers and detection models. Through a comparison of the outcomes produced by these models, valuable insights can be derived.

\begin{figure*}[tbp]
  \centering
  \includegraphics[width=\textwidth]{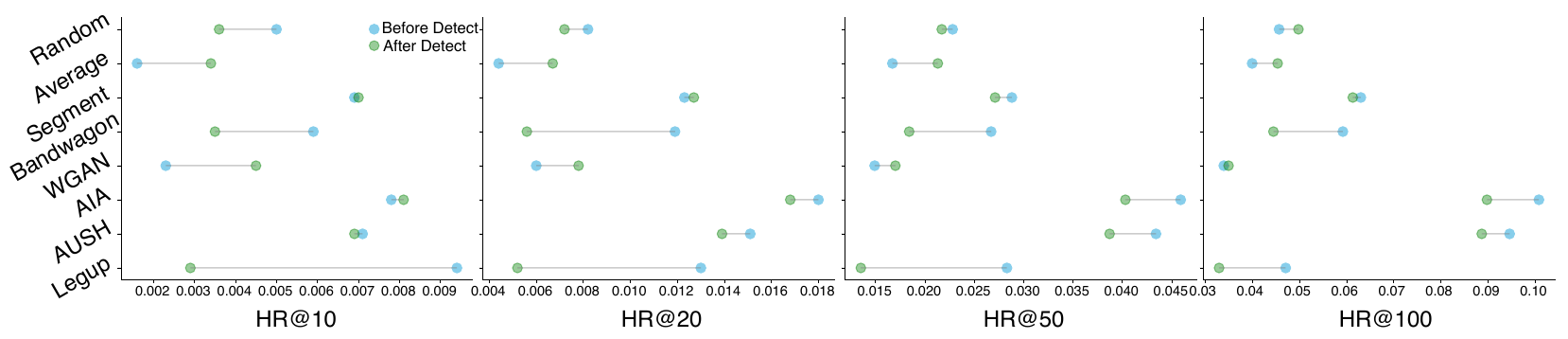}
  \caption{
  Performances of attackers before and after detection by PCASelectUser in the ML-1M dataset.
  }
  \label{fig:dot}
\end{figure*}

\begin{table*}[t]
    \renewcommand{\arraystretch}{1.5}
\setlength{\abovecaptionskip}{0cm}
\setlength{\belowcaptionskip}{0cm}
\caption{Defense performance against five representative shilling attackers.}
\label{tab:overall_per}
\begin{center}
\setlength{\tabcolsep}{0.2mm}{
\resizebox{\textwidth}{!}{
\begin{tabular}{@{}ccccccccccccccccc@{}}
\hline
\multicolumn{1}{l}{}                                   & \multicolumn{1}{l}{}                                   & \multicolumn{3}{c}{AIA}                                        & \multicolumn{3}{c}{Legup}                                      & \multicolumn{3}{c}{WGAN}                        & \multicolumn{3}{c}{RandomAttacker}              & \multicolumn{3}{c}{SegmentAttacker} \\
\multicolumn{1}{l}{}                                   & \multicolumn{1}{l}{}                                   & \multicolumn{3}{c}{Gray Box 20\% data} & \multicolumn{3}{c}{Gray Box 20\% data} & \multicolumn{3}{c}{Gray Box 20\% data}          & \multicolumn{3}{c}{White Box}                   & \multicolumn{3}{c}{White Box}       \\
Detect Method                                          & \multicolumn{1}{l}{Data Lable}                         & Precision      & Recall     & \multicolumn{1}{l}{F1-score}     & Precision      & Recall      & F1-score                        & Precision & Recall & F1-score                   & Precision & Recall & F1-score                   & Precision   & Recall   & F1-score   \\ \hline
\multicolumn{1}{c|}{DegreeSAD} & \multicolumn{1}{c|}{True Data} & 0.782          & 0.845      & \multicolumn{1}{c|}{0.812}       & 0.782          & 0.841       & \multicolumn{1}{c|}{0.810}      & 0.782     & 0.840  & \multicolumn{1}{c|}{0.810} & 0.780     & 0.843  & \multicolumn{1}{c|}{0.810} & 0.781       & 0.840    & 0.810      \\
\multicolumn{1}{l|}{}          & \multicolumn{1}{c|}{Fake Data} & 0.720          & 0.630      & \multicolumn{1}{c|}{0.672}       & 0.717          & 0.632       & \multicolumn{1}{c|}{0.672}      & 0.716     & 0.632  & \multicolumn{1}{c|}{0.671} & 0.718     & 0.627  & \multicolumn{1}{c|}{0.669} & 0.715       & 0.631    & 0.670      \\
\multicolumn{1}{c|}{CoDetector}                        & \multicolumn{1}{c|}{True Data} & 0.898          & 0.861      & \multicolumn{1}{c|}{0.879}       & 0.887          & 0.885       & \multicolumn{1}{c|}{0.886}      & 0.897     & 0.873  & \multicolumn{1}{c|}{0.885} & 0.908     & 0.877  & \multicolumn{1}{c|}{0.892} & 0.904       & 0.880    & 0.892      \\
\multicolumn{1}{l|}{}                                  & \multicolumn{1}{c|}{Fake Data} & 0.796          & 0.846      & \multicolumn{1}{c|}{0.820}       & 0.840          & 0.843       & \multicolumn{1}{c|}{0.841}      & 0.811     & 0.844  & \multicolumn{1}{c|}{0.827} & 0.809     & 0.854  & \multicolumn{1}{c|}{0.831} & 0.823       & 0.857    & 0.840      \\
\multicolumn{1}{c|}{BayesDetector}                     & \multicolumn{1}{c|}{True Data} & 0.943          & 0.946      & \multicolumn{1}{c|}{0.945}       & 0.945          & 0.945       & \multicolumn{1}{c|}{0.945}      & 0.936     & 0.943  & \multicolumn{1}{c|}{0.940} & 0.944     & 0.936  & \multicolumn{1}{c|}{0.940} & 0.938       & 0.943    & 0.940      \\
\multicolumn{1}{l|}{}                                  & \multicolumn{1}{c|}{Fake Data} & 0.915          & 0.910      & \multicolumn{1}{c|}{0.912}       & 0.914          & 0.913       & \multicolumn{1}{c|}{0.913}      & 0.909     & 0.899  & \multicolumn{1}{c|}{0.904} & 0.896     & 0.908  & \multicolumn{1}{c|}{0.902} & 0.909       & 0.902    & 0.905      \\
\multicolumn{1}{c|}{SemiSAD}                           & \multicolumn{1}{c|}{True Data} & 0.895          & 1.000      & \multicolumn{1}{c|}{0.945}       & 0.911          & 1.000       & \multicolumn{1}{c|}{0.954}      & 0.921     & 1.000  & \multicolumn{1}{c|}{0.959} & 0.903     & 1.000  & \multicolumn{1}{c|}{0.949} & 0.892       & 1.000    & 0.943      \\
\multicolumn{1}{l|}{}                                  & \multicolumn{1}{c|}{Fake Data} & 0.000          & 0.000      & \multicolumn{1}{c|}{0.000}       & 0.000          & 0.000       & \multicolumn{1}{c|}{0.000}      & 0.000     & 0.000  & \multicolumn{1}{c|}{0.000} & 0.000     & 0.000  & \multicolumn{1}{c|}{0.000} & 0.000       & 0.000    & 0.000      \\
\multicolumn{1}{c|}{PCASelectUser}                     & \multicolumn{1}{c|}{True Data} & 0.953          & 0.985      & \multicolumn{1}{c|}{0.969}       & 0.954          & 0.986       & \multicolumn{1}{c|}{0.970}      & 0.954     & 0.986  & \multicolumn{1}{c|}{0.970} & 0.952     & 0.983  & \multicolumn{1}{c|}{0.967} & 0.952       & 0.983    & 0.967      \\
\multicolumn{1}{l|}{}                                  & \multicolumn{1}{c|}{Fake Data} & 0.100          & 0.034      & \multicolumn{1}{c|}{0.050}       & 0.170          & 0.057       & \multicolumn{1}{c|}{0.086}      & 0.170     & 0.057  & \multicolumn{1}{c|}{0.086} & 0.000     & 0.000  & \multicolumn{1}{c|}{0.000} & 0.000       & 0.000    & 0.000      \\
\multicolumn{1}{c|}{FAP}                               & \multicolumn{1}{c|}{True Data} & 0.963          & 0.992      & \multicolumn{1}{c|}{0.977}       & 0.970          & 1.000       & \multicolumn{1}{c|}{0.985}      & 0.970     & 1.000  & \multicolumn{1}{c|}{0.985} & 0.872     & 0.296  & \multicolumn{1}{c|}{0.442} & 0.953       & 0.658    & 0.728      \\
\multicolumn{1}{c|}{}                                  & \multicolumn{1}{c|}{Fake Data} & 0.526          & 0.184      & \multicolumn{1}{c|}{0.272}       & 1.000          & 0.325       & \multicolumn{1}{c|}{0.491}      & 1.000     & 0.343  & \multicolumn{1}{c|}{0.511} & 0.920     & 0.647  & \multicolumn{1}{c|}{0.967} & 0.968       & 0.969    & 0.961      \\ \hline
\end{tabular}
}
}
\end{center}
\end{table*}

\subsection{Comparison of Attackers}
We illustrate the performance of all attackers in three recommendation datasets in Table \ref{tab:overall_per_1}. The goal of all attackers is to make the target items get higher rankings, \emph{i.e.}, larger HR@k. 

The two gray box methods, AIA and AUSH, exhibit the best performances across all metrics and datasets, which attests to the efficacy of neural network-based approaches. In contrast, the performance of Legup is less consistent. For instance, Legup displays optimal performance with respect to HR@10 in Amazon, whereas it experiences a decrease in rank, falling to the middle-lower range, with respect to HR@20, HR@50, and HR@100. Additionally, in Yelp, Legup performs inadequately across all metrics, and its weak robustness is further illustrated in Figure \ref{fig:dot}.
The Legup model has been observed to exhibit unstable performance, which can be attributed to its training methodology that involves the simultaneous use of three distinct models. This approach has resulted in the same training instability issues that are commonly associated with GANs. Specifically, the use of multiple models in training can lead to a lack of consistency in the learned representations across the different models. This, in turn, can create conflicts in the optimization process and cause the model's performance to become highly dependent on the initialization and training procedures. 

The heuristic method RandomAttacker exhibits the poorest performance across all metrics and datasets, even when compared to a situation in which no attacker is utilized. In other words, RandomAttacker not only fails to enhance the ranking of target items but also results in a lowered ranking for those items. Due to the highly randomized nature of heuristic attacks, the resulting attack target can be skewed by the random effects, resulting in a greater impact of inserting fake users on vulnerable users.

In addition, other methods, including SegmentAttack, BandwagonAttack, AverageAttack, and WGAN, also occasionally result in a poorer ranking for the targeted items. Consequently, there remains substantial room for the development of effective attacker methods in recommender systems. Currently, the existing methods of attack are characterized by significant limitations, such as their capacity to target only specific structures of recommendation algorithms or their limited ability to transfer attacks to models in other domains. 

\subsection{Comparison of Defenders}
We choose three supervised methods (DegreeSAD, CoDetector, and BayesDetector), one semi-supervised method (SemiSAD), and two unsupervised methods (PCASelectUser and FAP) to act as defenders, tasked with protecting the victim model from five attacker models. The goal of these experiments is to evaluate several defense methods using our framework process.

We present three evaluation metrics for the predicted label results, namely F1-score, Recall, and Precision. To evaluate the performance of our model, we split our data into two categories: True Data and Fake Data. True Data refers to the original real data used to train the recommender system, while Fake Data represents the fake data generated by attackers. We have provided three evaluation metrics for each category instead of treating them as a whole, as we believe that an effective detector should be able to not only successfully predict fake data but also avoid misclassifying real data. Hence, we hope that the values for the three metrics corresponding to both types of data are as high as possible, indicating that the detector models have better defensive performances from two dimensions.

Based on the data presented in Table \ref{tab:overall_per}, it can be observed that although the three supervised methods may not exhibit the highest performance, they demonstrate consistent performance against various attacks. Conversely, the semi-supervised method is not effective in defending against attacks due to the requirement of more data for training and evaluation, which is restricted by the attack budget in our approach. Consequently, the semi-supervised method misclassifies both real and fake data. Among the unsupervised methods, FAP shows promising results for certain attacks and outperforms other defense methods, but still displays certain limitations in some metrics.

\subsection{Robustness of Attackers Encountering Detection}
For illustration, we visualize the performance comparison before and after detection by PCASelectUser. Due to space limitations, we only present the results in the ML-1M dataset.

From Figure \ref{fig:dot}, we can observe that the performance of all attack methods will vary after detection. In heuristic methods, Bandwagon exhibits a notable difference in performance before and after detection. After detection, there is a marked decrease in all four HR metrics. The potential reason is that Bandwagon selects the popular items as users' fakes preferences, where this pattern is relatively easier to identify, making the generated data easier to be detected. 
In the neural methods, Legup demonstrates a similar phenomenon with a more significant performance difference before and after detection. In the HR@10 metric, Legup outperforms all other attacker methods before the detection, however, it has the worst performance after the detection. On HR@20, HR@50, and HR@100, it remains to be the worst one after detection. One possible reason for this is that Legup's optimization objective is more complex and it includes a greater number of modules compared to other attacker methods.
Both the two attackers show poor robustness before and after detection, while other methods exhibited relatively high robustness after detection. 

Counterintuitively, the results of AverageAttack and WGAN demonstrate an inverse effect: the targeted items rank higher after detection, \emph{i.e.}, the detection process helps the attackers achieve their purposes. There are two potential explanations. The first possibility is that this method generates users that are virtually indistinguishable from real ones, rendering detection modules theoretically unable to identify them. The second explanation is that the mechanism by which the method generates fake users was not taken into account by the detection module, allowing it to evade detection by this detection method.

\subsection{Comparison of Defense Evaluation. }
In light of the results presented in Figure \ref{fig:dot} and Table \ref{tab:overall_per}, we have observed that relying solely on either injection-based or label prediction evaluation for assessing the performance of defense models may not be adequate. For instance, in the case of the WGAN method, the injection-based evaluation indicates that the exposure rate of the target item is even higher after defense than the direct attack, while the label prediction evaluation suggests that the current defense approach is more effective in true and false prediction. Thus, we urge future researchers in this field to use both evaluation methods to ensure the practical effectiveness of defense models. Our framework supports both evaluation processes, eliminating the need for researchers to repeat work. 

\begin{figure}[]
\setlength{\belowcaptionskip}{-0.4cm}
\centering
\includegraphics[width=0.45\textwidth]{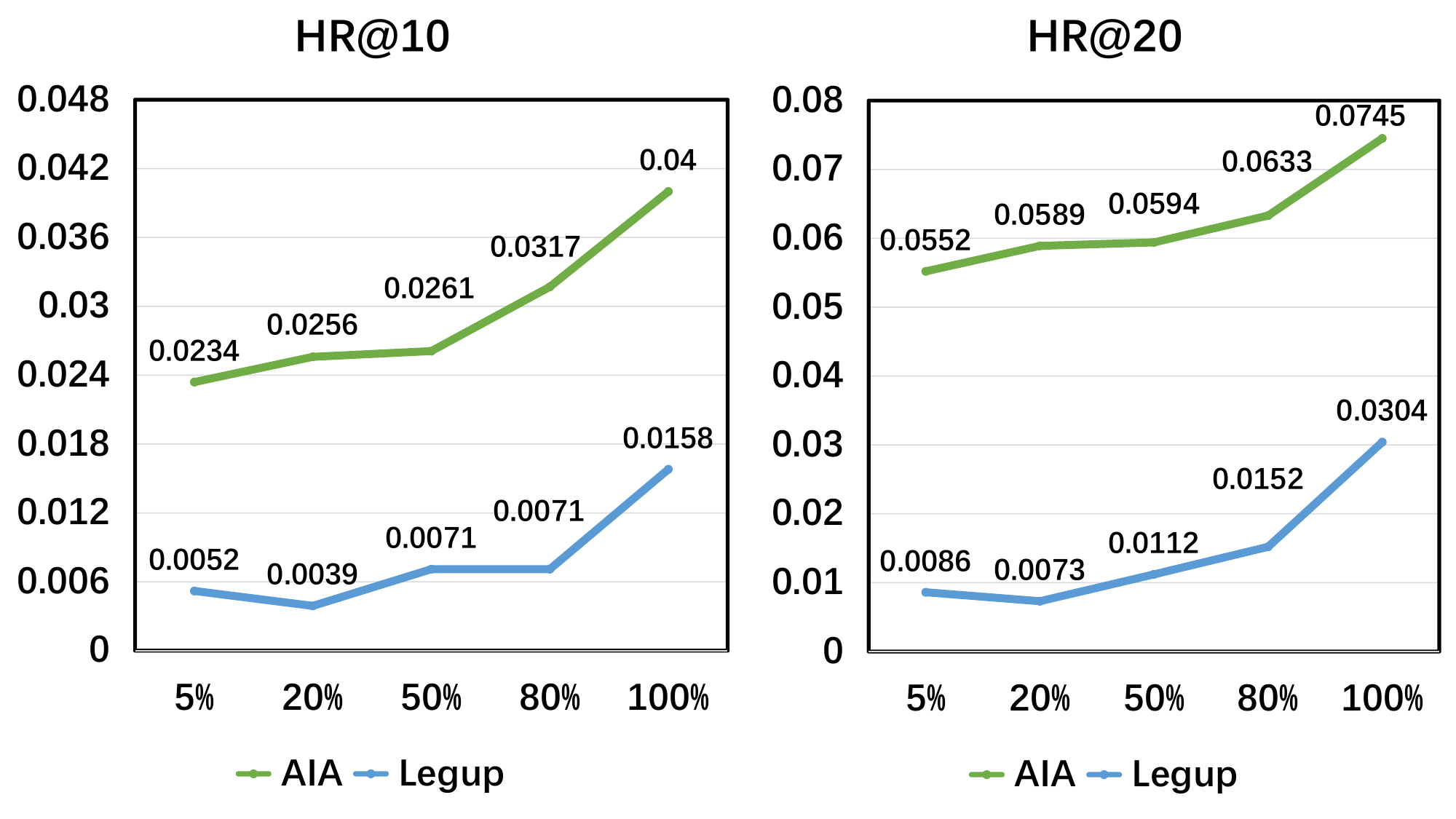}
\caption{Attack performance of AIA and Legup with different proportions of data.}
\label{fig:scale}
\end{figure}
\subsection{Effect of knowledge of the attackers.}
To investigate the impact of the scale of models' knowledge, \emph{i.e.}, the quantity of training data for attacker models, on the results, we visualize the performance of two neural models (AIA and Legup) as the amount of training data varies. The results are shown in Figure \ref{fig:scale}. From the results, we observe that the performance of both models increases as the amount of data increases, albeit with a slight fluctuation for AIA at x50\%. This inspires us to provide the attack model with more knowledge. However, as we venture into the exploration of novel attack algorithms, we must also take into consideration the importance of placing constraints on the known knowledge of these algorithms. This is particularly crucial, as it creates a trade-off between the scale of a model's knowledge and the effectiveness of the attack. Striking the right balance between these two factors is key to maximizing the potential impact of new attack algorithms while minimizing their potential negative consequences.
This trade-off raises the critical question of how we can best manage the scale of models' knowledge while still maintaining the efficacy of the attack. This requires a comprehensive understanding of the intricate interplay between the scale of knowledge and the effectiveness of the attack, and a willingness to explore new frontiers of research and development in order to push the boundaries of what is currently possible.

\section{Conclusion and future work}

Recommender systems have gained significant attention in recent years. However, the effectiveness and security of these systems have also become major concerns, as attackers may attempt to manipulate the recommendations for their own benefit. To promote research in this important field, we introduce RecAD, a new recommender library that provides a variety of benchmark datasets, evaluation settings, attackers, and defense models. 
By using RecAD, researchers can simulate a range of real-world scenarios and evaluate the robustness of different recommender systems against a variety of potential attacks.

In addition to advancing attacks and defenses on traditional models, we also acknowledge the transformative impact of large language models in the field of recommender systems \cite{wang2023generative,bao2023tallrec,zhang2023chatgpt}. Despite their powerful generative capabilities, these models are also susceptible to various attacks \cite{wang2023adversarial}. Therefore, our future research will also focus on the development of attack and defense mechanisms specifically tailored to large language model-based recommendations. In order to address this aim, We call upon researchers to collaborate and establish recommender system attack and defense methods that better align with the evolving needs of the field, enhancing the security and robustness of these models.

\bibliographystyle{ACM-Reference-Format}
\bibliography{sample-base}


\begin{thebibliography}{55}


\ifx \showCODEN    \undefined \def \showCODEN     #1{\unskip}     \fi
\ifx \showDOI      \undefined \def \showDOI       #1{#1}\fi
\ifx \showISBNx    \undefined \def \showISBNx     #1{\unskip}     \fi
\ifx \showISBNxiii \undefined \def \showISBNxiii  #1{\unskip}     \fi
\ifx \showISSN     \undefined \def \showISSN      #1{\unskip}     \fi
\ifx \showLCCN     \undefined \def \showLCCN      #1{\unskip}     \fi
\ifx \shownote     \undefined \def \shownote      #1{#1}          \fi
\ifx \showarticletitle \undefined \def \showarticletitle #1{#1}   \fi
\ifx \showURL      \undefined \def \showURL       {\relax}        \fi
\providecommand\bibfield[2]{#2}
\providecommand\bibinfo[2]{#2}
\providecommand\natexlab[1]{#1}
\providecommand\showeprint[2][]{arXiv:#2}

\bibitem[Aktukmak et~al\mbox{.}(2019)]%
        {aktukmak2019quick}
\bibfield{author}{\bibinfo{person}{Mehmet Aktukmak}, \bibinfo{person}{Yasin
  Yilmaz}, {and} \bibinfo{person}{Ismail Uysal}.}
  \bibinfo{year}{2019}\natexlab{}.
\newblock \showarticletitle{Quick and accurate attack detection in recommender
  systems through user attributes}. In \bibinfo{booktitle}{\emph{Proceedings of
  the 13th ACM Conference on Recommender Systems}}. \bibinfo{publisher}{ACM},
  \bibinfo{pages}{348--352}.
\newblock


\bibitem[Arjovsky et~al\mbox{.}(2017)]%
        {https://doi.org/10.48550/arxiv.1701.07875}
\bibfield{author}{\bibinfo{person}{Martin Arjovsky}, \bibinfo{person}{Soumith
  Chintala}, {and} \bibinfo{person}{L{\'e}on Bottou}.}
  \bibinfo{year}{2017}\natexlab{}.
\newblock \showarticletitle{{W}asserstein Generative Adversarial Networks}. In
  \bibinfo{booktitle}{\emph{Proceedings of the 34th International Conference on
  Machine Learning}}, \bibfield{editor}{\bibinfo{person}{Doina Precup} {and}
  \bibinfo{person}{Yee~Whye Teh}} (Eds.), Vol.~\bibinfo{volume}{70}.
  \bibinfo{publisher}{PMLR}, \bibinfo{pages}{214--223}.
\newblock


\bibitem[Bao et~al\mbox{.}(2023)]%
        {bao2023tallrec}
\bibfield{author}{\bibinfo{person}{Keqin Bao}, \bibinfo{person}{Jizhi Zhang},
  \bibinfo{person}{Yang Zhang}, \bibinfo{person}{Wenjie Wang},
  \bibinfo{person}{Fuli Feng}, {and} \bibinfo{person}{Xiangnan He}.}
  \bibinfo{year}{2023}\natexlab{}.
\newblock \bibinfo{title}{TALLRec: An Effective and Efficient Tuning Framework
  to Align Large Language Model with Recommendation}.
\newblock
\newblock


\bibitem[Bhaumik et~al\mbox{.}(2006)]%
        {bhaumik2006securing}
\bibfield{author}{\bibinfo{person}{Runa Bhaumik}, \bibinfo{person}{Chad
  Williams}, \bibinfo{person}{Bamshad Mobasher}, {and} \bibinfo{person}{Robin
  Burke}.} \bibinfo{year}{2006}\natexlab{}.
\newblock \showarticletitle{Securing collaborative filtering against malicious
  attacks through anomaly detection}. In \bibinfo{booktitle}{\emph{Proceedings
  of the 4th workshop on intelligent techniques for web personalization
  (ITWP’06), Boston}}. \bibinfo{publisher}{AAAI}, \bibinfo{pages}{10}.
\newblock


\bibitem[Burke et~al\mbox{.}(2005a)]%
        {burke2005limited}
\bibfield{author}{\bibinfo{person}{Robin Burke}, \bibinfo{person}{Bamshad
  Mobasher}, {and} \bibinfo{person}{Runa Bhaumik}.}
  \bibinfo{year}{2005}\natexlab{a}.
\newblock \showarticletitle{Limited knowledge shilling attacks in collaborative
  filtering systems}. In \bibinfo{booktitle}{\emph{Proceedings of 3rd
  international workshop on intelligent techniques for web personalization
  (ITWP), 19th international joint conference on artificial intelligence
  (IJCAI)}}. \bibinfo{publisher}{IJCAI}, \bibinfo{pages}{17--24}.
\newblock


\bibitem[Burke et~al\mbox{.}(2005b)]%
        {1565730}
\bibfield{author}{\bibinfo{person}{R. Burke}, \bibinfo{person}{B. Mobasher},
  \bibinfo{person}{R. Bhaumik}, {and} \bibinfo{person}{C. Williams}.}
  \bibinfo{year}{2005}\natexlab{b}.
\newblock \showarticletitle{Segment-based injection attacks against
  collaborative filtering recommender systems}. In
  \bibinfo{booktitle}{\emph{Fifth IEEE International Conference on Data Mining
  (ICDM'05)}}. \bibinfo{publisher}{IEEE}, \bibinfo{pages}{4 pp.--}.
\newblock


\bibitem[Cao et~al\mbox{.}(2013)]%
        {cao2013shilling}
\bibfield{author}{\bibinfo{person}{Jie Cao}, \bibinfo{person}{Zhiang Wu},
  \bibinfo{person}{Bo Mao}, {and} \bibinfo{person}{Yanchun Zhang}.}
  \bibinfo{year}{2013}\natexlab{}.
\newblock \showarticletitle{Shilling attack detection utilizing semi-supervised
  learning method for collaborative recommender system}.
\newblock \bibinfo{journal}{\emph{World Wide Web}}  \bibinfo{volume}{16}
  (\bibinfo{year}{2013}), \bibinfo{pages}{729--748}.
\newblock


\bibitem[Davoudi and Chatterjee(2017)]%
        {davoudi2017detection}
\bibfield{author}{\bibinfo{person}{Anahita Davoudi} {and}
  \bibinfo{person}{Mainak Chatterjee}.} \bibinfo{year}{2017}\natexlab{}.
\newblock \showarticletitle{Detection of profile injection attacks in social
  recommender systems using outlier analysis}. In
  \bibinfo{booktitle}{\emph{2017 IEEE International Conference on Big Data (Big
  Data)}}. \bibinfo{publisher}{IEEE}, \bibinfo{pages}{2714--2719}.
\newblock


\bibitem[Deldjoo et~al\mbox{.}(2019)]%
        {deldjoo2019assessing}
\bibfield{author}{\bibinfo{person}{Yashar Deldjoo}, \bibinfo{person}{Tommaso
  Di~Noia}, {and} \bibinfo{person}{Felice~Antonio Merra}.}
  \bibinfo{year}{2019}\natexlab{}.
\newblock \showarticletitle{Assessing the impact of a user-item collaborative
  attack on class of users}.
\newblock \bibinfo{journal}{\emph{arXiv preprint arXiv:1908.07968}}
  (\bibinfo{year}{2019}).
\newblock


\bibitem[Dou et~al\mbox{.}(2018)]%
        {dou2018collaborative}
\bibfield{author}{\bibinfo{person}{Tong Dou}, \bibinfo{person}{Junliang Yu},
  \bibinfo{person}{Qingyu Xiong}, \bibinfo{person}{Min Gao},
  \bibinfo{person}{Yuqi Song}, {and} \bibinfo{person}{Qianqi Fang}.}
  \bibinfo{year}{2018}\natexlab{}.
\newblock \showarticletitle{Collaborative shilling detection bridging
  factorization and user embedding}. In \bibinfo{booktitle}{\emph{Collaborative
  Computing: Networking, Applications and Worksharing}}.
  \bibinfo{publisher}{Springer}, \bibinfo{pages}{459--469}.
\newblock


\bibitem[Fan et~al\mbox{.}(2021)]%
        {fan2021attacking}
\bibfield{author}{\bibinfo{person}{Wenqi Fan}, \bibinfo{person}{Tyler Derr},
  \bibinfo{person}{Xiangyu Zhao}, \bibinfo{person}{Yao Ma},
  \bibinfo{person}{Hui Liu}, \bibinfo{person}{Jianping Wang},
  \bibinfo{person}{Jiliang Tang}, {and} \bibinfo{person}{Qing Li}.}
  \bibinfo{year}{2021}\natexlab{}.
\newblock \showarticletitle{Attacking black-box recommendations via copying
  cross-domain user profiles}. In \bibinfo{booktitle}{\emph{2021 IEEE 37th
  International Conference on Data Engineering (ICDE)}}.
  \bibinfo{publisher}{IEEE}, \bibinfo{pages}{1583--1594}.
\newblock


\bibitem[Fang et~al\mbox{.}(2020)]%
        {fang2020influence}
\bibfield{author}{\bibinfo{person}{Minghong Fang},
  \bibinfo{person}{Neil~Zhenqiang Gong}, {and} \bibinfo{person}{Jia Liu}.}
  \bibinfo{year}{2020}\natexlab{}.
\newblock \showarticletitle{Influence function based data poisoning attacks to
  top-n recommender systems}. In \bibinfo{booktitle}{\emph{Proceedings of The
  Web Conference}}. \bibinfo{publisher}{ACM}, \bibinfo{pages}{3019--3025}.
\newblock


\bibitem[Fang et~al\mbox{.}(2018a)]%
        {10.1145/3274694.3274706}
\bibfield{author}{\bibinfo{person}{Minghong Fang}, \bibinfo{person}{Guolei
  Yang}, \bibinfo{person}{Neil~Zhenqiang Gong}, {and} \bibinfo{person}{Jia
  Liu}.} \bibinfo{year}{2018}\natexlab{a}.
\newblock \showarticletitle{Poisoning Attacks to Graph-Based Recommender
  Systems}. In \bibinfo{booktitle}{\emph{Proceedings of the 34th Annual
  Computer Security Applications Conference}}. \bibinfo{publisher}{Association
  for Computing Machinery}, \bibinfo{address}{New York, NY, USA},
  \bibinfo{pages}{381–392}.
\newblock


\bibitem[Fang et~al\mbox{.}(2018b)]%
        {fang2018poisoning}
\bibfield{author}{\bibinfo{person}{Minghong Fang}, \bibinfo{person}{Guolei
  Yang}, \bibinfo{person}{Neil~Zhenqiang Gong}, {and} \bibinfo{person}{Jia
  Liu}.} \bibinfo{year}{2018}\natexlab{b}.
\newblock \showarticletitle{Poisoning attacks to graph-based recommender
  systems}. In \bibinfo{booktitle}{\emph{Proceedings of the 34th annual
  computer security applications conference}}. \bibinfo{publisher}{ACM},
  \bibinfo{pages}{381--392}.
\newblock


\bibitem[Gao et~al\mbox{.}(2022)]%
        {gao2022cirs}
\bibfield{author}{\bibinfo{person}{Chongming Gao}, \bibinfo{person}{Wenqiang
  Lei}, \bibinfo{person}{Jiawei Chen}, \bibinfo{person}{Shiqi Wang},
  \bibinfo{person}{Xiangnan He}, \bibinfo{person}{Shijun Li},
  \bibinfo{person}{Biao Li}, \bibinfo{person}{Yuan Zhang}, {and}
  \bibinfo{person}{Peng Jiang}.} \bibinfo{year}{2022}\natexlab{}.
\newblock \showarticletitle{Cirs: Bursting filter bubbles by counterfactual
  interactive recommender system}.
\newblock \bibinfo{journal}{\emph{arXiv preprint arXiv:2204.01266}}
  (\bibinfo{year}{2022}).
\newblock


\bibitem[Gao et~al\mbox{.}(2020)]%
        {gao2020shilling}
\bibfield{author}{\bibinfo{person}{Jianling Gao}, \bibinfo{person}{Lingtao Qi},
  \bibinfo{person}{Haiping Huang}, {and} \bibinfo{person}{Chao Sha}.}
  \bibinfo{year}{2020}\natexlab{}.
\newblock \showarticletitle{Shilling attack detection scheme in collaborative
  filtering recommendation system based on recurrent neural network}. In
  \bibinfo{booktitle}{\emph{Advances in Information and Communication:
  Proceedings of the 2020 Future of Information and Communication Conference
  (FICC), Volume 1}}. \bibinfo{publisher}{Springer}, \bibinfo{pages}{634--644}.
\newblock


\bibitem[Goodfellow et~al\mbox{.}(2020)]%
        {goodfellow2020generative}
\bibfield{author}{\bibinfo{person}{Ian Goodfellow}, \bibinfo{person}{Jean
  Pouget-Abadie}, \bibinfo{person}{Mehdi Mirza}, \bibinfo{person}{Bing Xu},
  \bibinfo{person}{David Warde-Farley}, \bibinfo{person}{Sherjil Ozair},
  \bibinfo{person}{Aaron Courville}, {and} \bibinfo{person}{Yoshua Bengio}.}
  \bibinfo{year}{2020}\natexlab{}.
\newblock \showarticletitle{Generative adversarial networks}.
\newblock \bibinfo{journal}{\emph{Commun. ACM}} \bibinfo{volume}{63},
  \bibinfo{number}{11} (\bibinfo{year}{2020}), \bibinfo{pages}{139--144}.
\newblock


\bibitem[He et~al\mbox{.}(2018)]%
        {he2018adversarial}
\bibfield{author}{\bibinfo{person}{Xiangnan He}, \bibinfo{person}{Zhankui He},
  \bibinfo{person}{Xiaoyu Du}, {and} \bibinfo{person}{Tat-Seng Chua}.}
  \bibinfo{year}{2018}\natexlab{}.
\newblock \showarticletitle{Adversarial personalized ranking for
  recommendation}. In \bibinfo{booktitle}{\emph{The 41st International ACM
  SIGIR conference on research \& development in information retrieval}}.
  \bibinfo{publisher}{ACM}, \bibinfo{pages}{355--364}.
\newblock


\bibitem[Huang et~al\mbox{.}(2021)]%
        {huang2021data}
\bibfield{author}{\bibinfo{person}{Hai Huang}, \bibinfo{person}{Jiaming Mu},
  \bibinfo{person}{Neil~Zhenqiang Gong}, \bibinfo{person}{Qi Li},
  \bibinfo{person}{Bin Liu}, {and} \bibinfo{person}{Mingwei Xu}.}
  \bibinfo{year}{2021}\natexlab{}.
\newblock \showarticletitle{Data poisoning attacks to deep learning based
  recommender systems}.
\newblock \bibinfo{journal}{\emph{arXiv preprint arXiv:2101.02644}}
  (\bibinfo{year}{2021}).
\newblock


\bibitem[Huang et~al\mbox{.}(2007)]%
        {4338497}
\bibfield{author}{\bibinfo{person}{Zan Huang}, \bibinfo{person}{Daniel Zeng},
  {and} \bibinfo{person}{Hsinchun Chen}.} \bibinfo{year}{2007}\natexlab{}.
\newblock \showarticletitle{A Comparison of Collaborative-Filtering
  Recommendation Algorithms for E-commerce}.
\newblock \bibinfo{journal}{\emph{IEEE Intelligent Systems}}
  \bibinfo{volume}{22} (\bibinfo{year}{2007}), \bibinfo{pages}{68--78}.
\newblock


\bibitem[Kaur and Goel(2016)]%
        {kaur2016shilling}
\bibfield{author}{\bibinfo{person}{Parneet Kaur} {and} \bibinfo{person}{Shivani
  Goel}.} \bibinfo{year}{2016}\natexlab{}.
\newblock \showarticletitle{Shilling attack models in recommender system}. In
  \bibinfo{booktitle}{\emph{ICICT}}, Vol.~\bibinfo{volume}{2}. IEEE,
  \bibinfo{pages}{1--5}.
\newblock


\bibitem[Lam and Riedl(2004)]%
        {lam2004shilling}
\bibfield{author}{\bibinfo{person}{Shyong~K Lam} {and} \bibinfo{person}{John
  Riedl}.} \bibinfo{year}{2004}\natexlab{}.
\newblock \showarticletitle{Shilling recommender systems for fun and profit}.
  In \bibinfo{booktitle}{\emph{Proceedings of the 13th international conference
  on World Wide Web}}. \bibinfo{publisher}{ACM}, \bibinfo{pages}{393--402}.
\newblock


\bibitem[Li et~al\mbox{.}(2016b)]%
        {li2016data}
\bibfield{author}{\bibinfo{person}{Bo Li}, \bibinfo{person}{Yining Wang},
  \bibinfo{person}{Aarti Singh}, {and} \bibinfo{person}{Yevgeniy Vorobeychik}.}
  \bibinfo{year}{2016}\natexlab{b}.
\newblock \showarticletitle{Data poisoning attacks on factorization-based
  collaborative filtering}.
\newblock \bibinfo{journal}{\emph{Advances in neural information processing
  systems}}  \bibinfo{volume}{29} (\bibinfo{year}{2016}).
\newblock


\bibitem[Li et~al\mbox{.}(2016a)]%
        {li2016shilling}
\bibfield{author}{\bibinfo{person}{Wentao Li}, \bibinfo{person}{Min Gao},
  \bibinfo{person}{Hua Li}, \bibinfo{person}{Jun Zeng}, \bibinfo{person}{Qingyu
  Xiong}, {and} \bibinfo{person}{Sachio Hirokawa}.}
  \bibinfo{year}{2016}\natexlab{a}.
\newblock \showarticletitle{Shilling attack detection in recommender systems
  via selecting patterns analysis}.
\newblock \bibinfo{journal}{\emph{IEICE TRANSACTIONS on Information and
  Systems}} \bibinfo{volume}{99}, \bibinfo{number}{10} (\bibinfo{year}{2016}),
  \bibinfo{pages}{2600--2611}.
\newblock


\bibitem[Lin et~al\mbox{.}(2020)]%
        {lin2020attacking}
\bibfield{author}{\bibinfo{person}{Chen Lin}, \bibinfo{person}{Si Chen},
  \bibinfo{person}{Hui Li}, \bibinfo{person}{Yanghua Xiao},
  \bibinfo{person}{Lianyun Li}, {and} \bibinfo{person}{Qian Yang}.}
  \bibinfo{year}{2020}\natexlab{}.
\newblock \showarticletitle{Attacking recommender systems with augmented user
  profiles}. In \bibinfo{booktitle}{\emph{Proceedings of the 29th ACM
  international conference on information \& knowledge management}}.
  \bibinfo{publisher}{ACM}, \bibinfo{pages}{855--864}.
\newblock


\bibitem[Lin et~al\mbox{.}(2022)]%
        {Lin_2022}
\bibfield{author}{\bibinfo{person}{Chen Lin}, \bibinfo{person}{Si Chen},
  \bibinfo{person}{Meifang Zeng}, \bibinfo{person}{Sheng Zhang},
  \bibinfo{person}{Min Gao}, {and} \bibinfo{person}{Hui Li}.}
  \bibinfo{year}{2022}\natexlab{}.
\newblock \showarticletitle{Shilling Black-Box Recommender Systems by Learning
  to Generate Fake User Profiles}.
\newblock \bibinfo{journal}{\emph{{IEEE} Transactions on Neural Networks and
  Learning Systems}} (\bibinfo{year}{2022}), \bibinfo{pages}{1--15}.
\newblock


\bibitem[Linden et~al\mbox{.}(2003)]%
        {1167344}
\bibfield{author}{\bibinfo{person}{G. Linden}, \bibinfo{person}{B. Smith},
  {and} \bibinfo{person}{J. York}.} \bibinfo{year}{2003}\natexlab{}.
\newblock \showarticletitle{Amazon.com recommendations: item-to-item
  collaborative filtering}.
\newblock \bibinfo{journal}{\emph{IEEE Internet Computing}}
  \bibinfo{volume}{7}, \bibinfo{number}{1} (\bibinfo{year}{2003}),
  \bibinfo{pages}{76--80}.
\newblock


\bibitem[Liu et~al\mbox{.}(2010)]%
        {10.1145/1719970.1719976}
\bibfield{author}{\bibinfo{person}{Jiahui Liu}, \bibinfo{person}{Peter Dolan},
  {and} \bibinfo{person}{Elin~R\o{}nby Pedersen}.}
  \bibinfo{year}{2010}\natexlab{}.
\newblock \showarticletitle{Personalized News Recommendation Based on Click
  Behavior}. In \bibinfo{booktitle}{\emph{Proceedings of the 15th International
  Conference on Intelligent User Interfaces}}. \bibinfo{publisher}{ACM},
  \bibinfo{pages}{31–40}.
\newblock


\bibitem[Liu et~al\mbox{.}(2020)]%
        {liu2020certifiable}
\bibfield{author}{\bibinfo{person}{Yang Liu}, \bibinfo{person}{Xianzhuo Xia},
  \bibinfo{person}{Liang Chen}, \bibinfo{person}{Xiangnan He},
  \bibinfo{person}{Carl Yang}, {and} \bibinfo{person}{Zibin Zheng}.}
  \bibinfo{year}{2020}\natexlab{}.
\newblock \showarticletitle{Certifiable robustness to discrete adversarial
  perturbations for factorization machines}. In
  \bibinfo{booktitle}{\emph{Proceedings of the 43rd International ACM SIGIR
  Conference on Research and Development in Information Retrieval}}.
  \bibinfo{publisher}{ACM}, \bibinfo{pages}{419--428}.
\newblock


\bibitem[Mehta and Nejdl(2009)]%
        {mehta2009unsupervised}
\bibfield{author}{\bibinfo{person}{Bhaskar Mehta} {and}
  \bibinfo{person}{Wolfgang Nejdl}.} \bibinfo{year}{2009}\natexlab{}.
\newblock \showarticletitle{Unsupervised strategies for shilling detection and
  robust collaborative filtering}.
\newblock \bibinfo{journal}{\emph{User Modeling and User-Adapted Interaction}}
  \bibinfo{volume}{19} (\bibinfo{year}{2009}), \bibinfo{pages}{65--97}.
\newblock


\bibitem[Mobasher et~al\mbox{.}(2007)]%
        {mobasher2007toward}
\bibfield{author}{\bibinfo{person}{Bamshad Mobasher}, \bibinfo{person}{Robin
  Burke}, \bibinfo{person}{Runa Bhaumik}, {and} \bibinfo{person}{Chad
  Williams}.} \bibinfo{year}{2007}\natexlab{}.
\newblock \showarticletitle{Toward trustworthy recommender systems: An analysis
  of attack models and algorithm robustness}.
\newblock \bibinfo{journal}{\emph{ACM Transactions on Internet Technology
  (TOIT)}} \bibinfo{volume}{7}, \bibinfo{number}{4} (\bibinfo{year}{2007}),
  \bibinfo{pages}{23--es}.
\newblock


\bibitem[O'Mahony et~al\mbox{.}(2005)]%
        {o2005recommender}
\bibfield{author}{\bibinfo{person}{Michael~P O'Mahony}, \bibinfo{person}{Neil~J
  Hurley}, {and} \bibinfo{person}{Gu{\'e}nol{\'e}~CM Silvestre}.}
  \bibinfo{year}{2005}\natexlab{}.
\newblock \showarticletitle{Recommender systems: Attack types and strategies}.
  In \bibinfo{booktitle}{\emph{Association for the Advancement of Artificial
  Intelligence (AAAI)}}. \bibinfo{publisher}{AAAI}, \bibinfo{pages}{334--339}.
\newblock


\bibitem[Rong et~al\mbox{.}(2022a)]%
        {rong2022poisoning}
\bibfield{author}{\bibinfo{person}{Dazhong Rong}, \bibinfo{person}{Qinming He},
  {and} \bibinfo{person}{Jianhai Chen}.} \bibinfo{year}{2022}\natexlab{a}.
\newblock \showarticletitle{Poisoning Deep Learning based Recommender Model in
  Federated Learning Scenarios}.
\newblock \bibinfo{journal}{\emph{arXiv preprint arXiv:2204.13594}}
  (\bibinfo{year}{2022}).
\newblock


\bibitem[Rong et~al\mbox{.}(2022b)]%
        {rong2022fedrecattack}
\bibfield{author}{\bibinfo{person}{Dazhong Rong}, \bibinfo{person}{Shuai Ye},
  \bibinfo{person}{Ruoyan Zhao}, \bibinfo{person}{Hon~Ning Yuen},
  \bibinfo{person}{Jianhai Chen}, {and} \bibinfo{person}{Qinming He}.}
  \bibinfo{year}{2022}\natexlab{b}.
\newblock \showarticletitle{Fedrecattack: Model poisoning attack to federated
  recommendation}. In \bibinfo{booktitle}{\emph{2022 IEEE 38th International
  Conference on Data Engineering (ICDE)}}. \bibinfo{publisher}{IEEE},
  \bibinfo{pages}{2643--2655}.
\newblock


\bibitem[Shahrasbi et~al\mbox{.}(2020)]%
        {DBLP:journals/corr/abs-2012-02509}
\bibfield{author}{\bibinfo{person}{Behzad Shahrasbi},
  \bibinfo{person}{Venugopal Mani}, \bibinfo{person}{Apoorv~Reddy Arrabothu},
  \bibinfo{person}{Deepthi Sharma}, \bibinfo{person}{Kannan Achan}, {and}
  \bibinfo{person}{Sushant Kumar}.} \bibinfo{year}{2020}\natexlab{}.
\newblock \showarticletitle{On Detecting Data Pollution Attacks On Recommender
  Systems Using Sequential GANs}.
\newblock \bibinfo{journal}{\emph{CoRR}}  \bibinfo{volume}{abs/2012.02509}
  (\bibinfo{year}{2020}).
\newblock


\bibitem[Song et~al\mbox{.}(2020)]%
        {song2020poisonrec}
\bibfield{author}{\bibinfo{person}{Junshuai Song}, \bibinfo{person}{Zhao Li},
  \bibinfo{person}{Zehong Hu}, \bibinfo{person}{Yucheng Wu},
  \bibinfo{person}{Zhenpeng Li}, \bibinfo{person}{Jian Li}, {and}
  \bibinfo{person}{Jun Gao}.} \bibinfo{year}{2020}\natexlab{}.
\newblock \showarticletitle{Poisonrec: an adaptive data poisoning framework for
  attacking black-box recommender systems}. In \bibinfo{booktitle}{\emph{2020
  IEEE 36th International Conference on Data Engineering (ICDE)}}. IEEE,
  \bibinfo{pages}{157--168}.
\newblock


\bibitem[Tang et~al\mbox{.}(2019)]%
        {tang2019adversarial}
\bibfield{author}{\bibinfo{person}{Jinhui Tang}, \bibinfo{person}{Xiaoyu Du},
  \bibinfo{person}{Xiangnan He}, \bibinfo{person}{Fajie Yuan},
  \bibinfo{person}{Qi Tian}, {and} \bibinfo{person}{Tat-Seng Chua}.}
  \bibinfo{year}{2019}\natexlab{}.
\newblock \showarticletitle{Adversarial training towards robust multimedia
  recommender system}.
\newblock \bibinfo{journal}{\emph{IEEE Transactions on Knowledge and Data
  Engineering}} \bibinfo{volume}{32}, \bibinfo{number}{5}
  (\bibinfo{year}{2019}), \bibinfo{pages}{855--867}.
\newblock


\bibitem[Tang et~al\mbox{.}(2020)]%
        {tang2020revisiting}
\bibfield{author}{\bibinfo{person}{Jiaxi Tang}, \bibinfo{person}{Hongyi Wen},
  {and} \bibinfo{person}{Ke Wang}.} \bibinfo{year}{2020}\natexlab{}.
\newblock \showarticletitle{Revisiting adversarially learned injection attacks
  against recommender systems}. In \bibinfo{booktitle}{\emph{Proceedings of the
  14th ACM Conference on Recommender Systems}}. \bibinfo{publisher}{ACM},
  \bibinfo{pages}{318--327}.
\newblock


\bibitem[Wang et~al\mbox{.}(2023b)]%
        {wang2023adversarial}
\bibfield{author}{\bibinfo{person}{Jiongxiao Wang}, \bibinfo{person}{Zichen
  Liu}, \bibinfo{person}{Keun~Hee Park}, \bibinfo{person}{Muhao Chen}, {and}
  \bibinfo{person}{Chaowei Xiao}.} \bibinfo{year}{2023}\natexlab{b}.
\newblock \bibinfo{title}{Adversarial Demonstration Attacks on Large Language
  Models}.
\newblock
\newblock


\bibitem[Wang et~al\mbox{.}(2023a)]%
        {wang2023generative}
\bibfield{author}{\bibinfo{person}{Wenjie Wang}, \bibinfo{person}{Xinyu Lin},
  \bibinfo{person}{Fuli Feng}, \bibinfo{person}{Xiangnan He}, {and}
  \bibinfo{person}{Tat-Seng Chua}.} \bibinfo{year}{2023}\natexlab{a}.
\newblock \bibinfo{title}{Generative Recommendation: Towards Next-generation
  Recommender Paradigm}.
\newblock
\newblock


\bibitem[Wu et~al\mbox{.}(2021)]%
        {wu2021fight}
\bibfield{author}{\bibinfo{person}{Chenwang Wu}, \bibinfo{person}{Defu Lian},
  \bibinfo{person}{Yong Ge}, \bibinfo{person}{Zhihao Zhu},
  \bibinfo{person}{Enhong Chen}, {and} \bibinfo{person}{Senchao Yuan}.}
  \bibinfo{year}{2021}\natexlab{}.
\newblock \showarticletitle{Fight fire with fire: towards robust recommender
  systems via adversarial poisoning training}. In
  \bibinfo{booktitle}{\emph{Proceedings of the 44th International ACM SIGIR
  Conference on Research and Development in Information Retrieval}}.
  \bibinfo{publisher}{ACM}, \bibinfo{pages}{1074--1083}.
\newblock


\bibitem[Wu et~al\mbox{.}(2011)]%
        {wu2011semi}
\bibfield{author}{\bibinfo{person}{Zhiang Wu}, \bibinfo{person}{Jie Cao},
  \bibinfo{person}{Bo Mao}, {and} \bibinfo{person}{Youquan Wang}.}
  \bibinfo{year}{2011}\natexlab{}.
\newblock \showarticletitle{Semi-SAD: applying semi-supervised learning to
  shilling attack detection}. In \bibinfo{booktitle}{\emph{Proceedings of the
  fifth ACM conference on Recommender systems}}. \bibinfo{publisher}{ACM},
  \bibinfo{pages}{289--292}.
\newblock


\bibitem[Xing et~al\mbox{.}(2013)]%
        {182952}
\bibfield{author}{\bibinfo{person}{Xingyu Xing}, \bibinfo{person}{Wei Meng},
  \bibinfo{person}{Dan Doozan}, \bibinfo{person}{Alex~C. Snoeren},
  \bibinfo{person}{Nick Feamster}, {and} \bibinfo{person}{Wenke Lee}.}
  \bibinfo{year}{2013}\natexlab{}.
\newblock \showarticletitle{Take This Personally: Pollution Attacks on
  Personalized Services}. In \bibinfo{booktitle}{\emph{22nd USENIX Security
  Symposium (USENIX Security 13)}}. \bibinfo{publisher}{USENIX Association},
  \bibinfo{pages}{671--686}.
\newblock


\bibitem[Yang et~al\mbox{.}(2018)]%
        {yang2018detection}
\bibfield{author}{\bibinfo{person}{Fan Yang}, \bibinfo{person}{Min Gao},
  \bibinfo{person}{Junliang Yu}, \bibinfo{person}{Yuqi Song}, {and}
  \bibinfo{person}{Xinyi Wang}.} \bibinfo{year}{2018}\natexlab{}.
\newblock \showarticletitle{Detection of shilling attack based on bayesian
  model and user embedding}. In \bibinfo{booktitle}{\emph{2018 IEEE 30th
  International Conference on Tools with Artificial Intelligence (ICTAI)}}.
  \bibinfo{publisher}{IEEE}, \bibinfo{pages}{639--646}.
\newblock


\bibitem[Yang et~al\mbox{.}(2017)]%
        {yang2017fake}
\bibfield{author}{\bibinfo{person}{Guolei Yang},
  \bibinfo{person}{Neil~Zhenqiang Gong}, {and} \bibinfo{person}{Ying Cai}.}
  \bibinfo{year}{2017}\natexlab{}.
\newblock \showarticletitle{Fake Co-visitation Injection Attacks to Recommender
  Systems}. In \bibinfo{booktitle}{\emph{Network and Distributed System
  Security Symposium}}.
\newblock


\bibitem[Yi et~al\mbox{.}(2022)]%
        {https://doi.org/10.48550/arxiv.2202.06701}
\bibfield{author}{\bibinfo{person}{Jingwei Yi}, \bibinfo{person}{Fangzhao Wu},
  \bibinfo{person}{Bin Zhu}, \bibinfo{person}{Yang Yu}, \bibinfo{person}{Chao
  Zhang}, \bibinfo{person}{Guangzhong Sun}, {and} \bibinfo{person}{Xing Xie}.}
  \bibinfo{year}{2022}\natexlab{}.
\newblock \showarticletitle{UA-FedRec: Untargeted Attack on Federated News
  Recommendation}.
\newblock \bibinfo{journal}{\emph{CoRR}}  \bibinfo{volume}{abs/2202.06701}
  (\bibinfo{year}{2022}).
\newblock


\bibitem[Yu et~al\mbox{.}(2021)]%
        {9498095}
\bibfield{author}{\bibinfo{person}{Hongtao Yu}, \bibinfo{person}{Haihong
  Zheng}, \bibinfo{person}{Yishu Xu}, \bibinfo{person}{Ru Ma},
  \bibinfo{person}{Dingli Gao}, {and} \bibinfo{person}{Fuzhi Zhang}.}
  \bibinfo{year}{2021}\natexlab{}.
\newblock \showarticletitle{Detecting group shilling attacks in recommender
  systems based on maximum dense subtensor mining}. In
  \bibinfo{booktitle}{\emph{2021 IEEE International Conference on Artificial
  Intelligence and Computer Applications}}. \bibinfo{publisher}{IEEE},
  \bibinfo{pages}{644--648}.
\newblock


\bibitem[Zeller and Felten(2008)]%
        {zeller2008cross}
\bibfield{author}{\bibinfo{person}{William Zeller} {and}
  \bibinfo{person}{Edward~W Felten}.} \bibinfo{year}{2008}\natexlab{}.
\newblock \showarticletitle{Cross-site request forgeries: Exploitation and
  prevention}.
\newblock \bibinfo{journal}{\emph{The New York Times}} (\bibinfo{year}{2008}),
  \bibinfo{pages}{1--13}.
\newblock


\bibitem[Zhang et~al\mbox{.}(2020)]%
        {zhang2020practical}
\bibfield{author}{\bibinfo{person}{Hengtong Zhang}, \bibinfo{person}{Yaliang
  Li}, \bibinfo{person}{Bolin Ding}, {and} \bibinfo{person}{Jing Gao}.}
  \bibinfo{year}{2020}\natexlab{}.
\newblock \showarticletitle{Practical data poisoning attack against next-item
  recommendation}. In \bibinfo{booktitle}{\emph{Proceedings of The Web
  Conference}}. \bibinfo{publisher}{ACM}, \bibinfo{pages}{2458--2464}.
\newblock


\bibitem[Zhang et~al\mbox{.}(2023)]%
        {zhang2023chatgpt}
\bibfield{author}{\bibinfo{person}{Jizhi Zhang}, \bibinfo{person}{Keqin Bao},
  \bibinfo{person}{Yang Zhang}, \bibinfo{person}{Wenjie Wang},
  \bibinfo{person}{Fuli Feng}, {and} \bibinfo{person}{Xiangnan He}.}
  \bibinfo{year}{2023}\natexlab{}.
\newblock \bibinfo{title}{Is ChatGPT Fair for Recommendation? Evaluating
  Fairness in Large Language Model Recommendation}.
\newblock
\newblock


\bibitem[Zhang et~al\mbox{.}(2021)]%
        {10.1145/3442381.3449813}
\bibfield{author}{\bibinfo{person}{Shijie Zhang}, \bibinfo{person}{Hongzhi
  Yin}, \bibinfo{person}{Tong Chen}, \bibinfo{person}{Zi Huang},
  \bibinfo{person}{Lizhen Cui}, {and} \bibinfo{person}{Xiangliang Zhang}.}
  \bibinfo{year}{2021}\natexlab{}.
\newblock \showarticletitle{Graph Embedding for Recommendation against
  Attribute Inference Attacks}. In \bibinfo{booktitle}{\emph{Proceedings of the
  Web Conference 2021}}. \bibinfo{publisher}{ACM},
  \bibinfo{pages}{3002–3014}.
\newblock


\bibitem[Zhang et~al\mbox{.}(2022)]%
        {zhang2022targeted}
\bibfield{author}{\bibinfo{person}{Xudong Zhang}, \bibinfo{person}{Zan Wang},
  \bibinfo{person}{Jingke Zhao}, {and} \bibinfo{person}{Lanjun Wang}.}
  \bibinfo{year}{2022}\natexlab{}.
\newblock \showarticletitle{Targeted Data Poisoning Attack on News
  Recommendation System}.
\newblock \bibinfo{journal}{\emph{arXiv preprint arXiv:2203.03560}}
  (\bibinfo{year}{2022}).
\newblock


\bibitem[Zhao et~al\mbox{.}(2021)]%
        {recbole[1.0]}
\bibfield{author}{\bibinfo{person}{Wayne~Xin Zhao}, \bibinfo{person}{Shanlei
  Mu}, \bibinfo{person}{Yupeng Hou}, \bibinfo{person}{Zihan Lin},
  \bibinfo{person}{Yushuo Chen}, \bibinfo{person}{Xingyu Pan},
  \bibinfo{person}{Kaiyuan Li}, \bibinfo{person}{Yujie Lu},
  \bibinfo{person}{Hui Wang}, \bibinfo{person}{Changxin Tian},
  \bibinfo{person}{Yingqian Min}, \bibinfo{person}{Zhichao Feng},
  \bibinfo{person}{Xinyan Fan}, \bibinfo{person}{Xu Chen},
  \bibinfo{person}{Pengfei Wang}, \bibinfo{person}{Wendi Ji},
  \bibinfo{person}{Yaliang Li}, \bibinfo{person}{Xiaoling Wang}, {and}
  \bibinfo{person}{Ji{-}Rong Wen}.} \bibinfo{year}{2021}\natexlab{}.
\newblock \showarticletitle{RecBole: Towards a Unified, Comprehensive and
  Efficient Framework for Recommendation Algorithms}. In
  \bibinfo{booktitle}{\emph{{CIKM}}}. \bibinfo{publisher}{{ACM}},
  \bibinfo{pages}{4653--4664}.
\newblock


\bibitem[Zhou et~al\mbox{.}(2016)]%
        {zhou2016svm}
\bibfield{author}{\bibinfo{person}{Wei Zhou}, \bibinfo{person}{Junhao Wen},
  \bibinfo{person}{Qingyu Xiong}, \bibinfo{person}{Min Gao}, {and}
  \bibinfo{person}{Jun Zeng}.} \bibinfo{year}{2016}\natexlab{}.
\newblock \showarticletitle{SVM-TIA a shilling attack detection method based on
  SVM and target item analysis in recommender systems}.
\newblock \bibinfo{journal}{\emph{Neurocomputing}}  \bibinfo{volume}{210}
  (\bibinfo{year}{2016}), \bibinfo{pages}{197--205}.
\newblock


\bibitem[Zhou et~al\mbox{.}(2021)]%
        {zhou2021deep}
\bibfield{author}{\bibinfo{person}{Xingchen Zhou}, \bibinfo{person}{Ming Xu},
  \bibinfo{person}{Yiming Wu}, {and} \bibinfo{person}{Ning Zheng}.}
  \bibinfo{year}{2021}\natexlab{}.
\newblock \showarticletitle{Deep model poisoning attack on federated learning}.
\newblock \bibinfo{journal}{\emph{Future Internet}} \bibinfo{volume}{13},
  \bibinfo{number}{3} (\bibinfo{year}{2021}), \bibinfo{pages}{73}.
\newblock


\end{thebibliography}

\end{document}